\documentclass[12pt]{article}
\begin{document}
\title{ Evidence for $M_B$ and $M_C$ phases in the morphotropic phase boundary
region of $(1-x)[Pb(Mg_{1/3}Nb_{2/3})O_3]-xPbTiO_3$ : A Rietveld study}
\author{Akhilesh Kumar Singh and Dhananjai Pandey \thanks{email:dpandey@banras.ernet.in}\\ 
School of Materials Science \& Technology
\\ Banaras Hindu University, Varanasi 221 005, India}

\maketitle
\begin{abstract}
We present here the results of the room temperature dielectric constant measurements and Rietveld 
analysis of the powder x-ray diffraction data on $(1-x)[Pb(Mg_{1/3}Nb_{2/3})O_3]-xPbTiO_3$(PMN-$x$PT)
in the composition range $0.20 \leq x \leq 0.45$ 
to show that the morphotropic phase boundary (MPB) region contains two monoclinic phases 
with space groups Cm (or $M_B$ type) and Pm (or $M_C$ type) stable in the composition ranges 
$0.27 \leq x \leq 0.30$ and  $0.31 \leq x \leq 0.34$, respectively.
The structure of PMN-$x$PT in the composition ranges $0 \leq x \leq$ 0.26, 
and $0.35 \leq x \leq1$  is found to be rhombohedral (R3m) and 
tetragonal (P4mm), respectively. These results are compared with the predictions of  
Vanderbilt \& Cohen's theory. 
\end{abstract}                                                              
\section{Introduction}

Relaxor ferroelectric based morphotropic phase boundary (MPB) ceramics like 
$(1-x)[Pb({Mg_{1/3}} {Nb_{2/3}})O_3]-xPbTiO_3 (PMN-xPT)$ \cite{1} and 
$(1-x)[Pb(Zn_{1/3}Nb_{2/3})O_3]-xPbTiO_3 (PZN-xPT)$ \cite{2} show much higher electromechanical 
response about their MPBs in comparison to the well known $Pb(Zr_xTi_{1-x})O_3$ (PZT) system 
\cite{3}. The reason why relaxor based MPB systems have much higher electromechanical 
response is still not very clear eventhough recent theoretical and experimental developments in 
PZT have improved our understanding of the phase stabilities in the vicinity of the MPB 
\cite {4,5,6,7,8,9,10,11,12}.  Noheda et al have discovered that the tetragonal phase 
of PZT with compositions ($x=0.50, 0.52$) close to the MPB transforms to a monoclinic phase
 with space group Cm at low temperatures \cite{4,5}. Ragini  et al  \cite{6} and 
Ranjan et al  \cite{7} have discovered yet another low temperature phase transition 
in which the Cm monoclinic phase transforms to another monoclinic phase with Cc space 
group \cite{8}. The Cm to Cc phase transition is an antiferrodistortive (AFD) transition 
leading to superlattice reflections which are observable in the electron \cite{6} and neutron \cite{7} 
diffraction patterns only and not in the XRD patterns as a result of which Noheda et al missed 
the Cc phase in their high resolution XRD studies at low temperatures. The tetragonal to Cm, 
and Cm to Cc transitions are accompanied with pronounced anomalies in the elastic constant 
and dielectric constant \cite{6}. It has been argued \cite{9,10,11} that the monoclinic Cm phase provides 
the path for polarization rotation between tetragonal (P4mm) and rhombohedral (R3m) phases.
 X-ray Rietveld analysis by Ragini et al \cite{12}, has, however, revealed that the hitherto believed
 rhombohedral phase of PZT for $0.53 < x < 0.62$ \cite{13,14} is indeed monoclinic (Cm) with very 
small domain size leading to composition dependent anomalous broadening of various reflections. 
Thus the MPB in the PZT system separates the stability fields of tetragonal and monoclinic 
Cm phases \cite{12}. On application of DC field, these monoclinic domains get aligned and merged 
as a result of which some of the XRD peaks showing anomalous broadening in the unpoled 
state start exhibiting splittings characteristic of the Cm phase in the poled state. Thus, according 
to Ragini et al \cite{12,15}, the field induced rhombohedral to Cm transition reported by 
Guo et al \cite{11} is really a transition from a small-domain Cm phase to large-domain Cm phase. 
Based on Rietveld analysis of the XRD data, Ragini et al \cite{12} have also shown that the tetragonal and 
monoclinic (Cm) phases coexist across the MPB in the composition range $0.520 \leq x \leq  0.525$ due to   
nucleation barrier to the first order transition between the high temperature tetragonal 
and the low temperature monoclinic Cm phases \cite{15}. 

These recent developments in PZT have been followed up by similar studies on the structure of MPB 
phases in the PMN-$x$PT \cite{16,17,18,19,20} and PZN-$x$PT \cite{21,22} systems. In these systems also, 
the structure of the morphotropic phase in the unpoled state is monoclinic but with a space group 
Pm \cite{16,17}, which is different from that in the PZT system. The Pm space group proposed by 
Singh and Pandey \cite{16} and Kiat et al \cite{17} in the MPB region of the PMN-xPT system has recently 
been confirmed in the high resolution powder XRD studies by Noheda et al \cite{23} contradicting an earlier 
report of Cm 
space group by the same workers for a similar composition \cite{21}. However in the PZN-$x$PT system, 
the possibility of monoclinic Pm \cite{17} or orthorhombic Bmm2 \cite{21} structures in the MPB region 
continues to be debated. More interestingly, Noheda et al \cite{24} and Ohwada, et al \cite{25} have observed 
a field-induced 
irreversible rhombohedral to Pm monoclinic phase transition in PZN-0.8PT through the Cm 
monoclinic phase. 

Using eighth-order expansion of Devonshire theory, Vanderbilt and Cohen
\cite{26} have predicted different regions of stability for the monoclinic Pm and Cm phases that are 
designated as $M_C$ and $M_A/M_B$, respectively, in their paper. Although both $M_A$ and $M_B$ phases 
belong to the Cm space group, the difference lies in the magnitudes of the components of the 
polarization ($P$) corresponding to the pseudocubic cell. For the $M_A$ phase $P_X = P_Y \neq P_Z$ with 
$P_Z >P_X$, while for the $M_B$ phase, $P_X = P_Y \neq P_Z$ with $P_Z < Px$. As per the phase diagram 
of Vanderbilt and Cohen \cite{26}, one expects a narrow stability field of the $M_B$ phase between the 
rhombohedral and $M_C$ phases. The present work was undertaken to verify the existence of the Cm 
($M_B$) phase for compositions in between those for the rhombohedral and monoclinic (Pm) phases 
using Rietveld analysis of X-ray powder diffraction data on PMN-$x$PT samples with $x$ varying 
from 0.26 to 0.39 at an interval of $\Delta x$ = 0.01. From a careful study of the variation of the dielectric 
constant and crystal structure with composition ($x$) on unpoled PMN-$x$PT ceramics, we show that the 
dominant phases in the composition ranges $0 \leq x \leq$ 0.26, $0.27 \leq x \leq 0.30$, $0.31 \leq x \leq 0.34$
 and $0.35 \leq x \leq1$  are rhombohedral (R3m), monoclinic $M_B$ (Cm), monoclinic $M_C$ 
(Pm) and tetragonal (P4mm), respectively, in reasonable agreement with the predictions of 
the Vanderbilt and Cohen's theory \cite{26}. 

\section{Experimental}

Samples used in the present work were prepared by a modified solid state route \cite{27}. One of the 
common problems associated with the solid state synthesis of PMN-$x$PT ceramics is the 
appearance of an unwanted pyrochlore phase \cite{28}. To get rid of this unwanted phase, 
off-stoichiometric compositions, with excess of MgO and PbO, are used \cite{29}. For example, 
Noheda et al \cite{23} have used 15.5 and 2wt$ \%$ excess of MgO and PbO for getting pure 
perovskite phase. This naturally perturbs the phase stabilities in the vicinity of the MPB where 
the crystal structure is very sensitive to even small variations in the composition because of 
nearly degenerate nature of various phases. In order to bring out the intrinsic features of the 
PMN-$x$PT system, it is imperative to prepare pyrochlore phase free PMN-$x$PT ceramics in 
stoichiometric compositions (i.e., without using any excess of PbO and MgO). We have achieved 
this by using $PbCO_3$ and $MgCO_3.3H_2O$, instead of PbO and MgO, respectively, and introducing 
one more step in the reaction sequence for mixing of $TiO_2$ \cite{27}. In the present work, AR grade 
$Nb_2O_5$ (99.95 \% ) , $TiO_2$ (99 \%), $Mg(NO_3)_2.6H_2O$ (99\%), $Pb(NO_3)_2$ 
(99 \%)  and ammonium carbonate were used. $MgCO_3.3H_2O$ and $PbCO_3$ were prepared 
from $Mg(NO_3)_2.6H_2O$ and $Pb(NO_3)_2$ by precipitation. Mixing of various ingredients 
in stoichiometric proportions was carried out for 
6 hours using a ball mill (Retsch, Japan) with zirconia jars and zirconia balls. AR grade acetone 
was used as the mixing media. Heat treatments for calcination were carried out in alumina 
crucibles using a globar furnace. The columbite precursor $MgNb_2O_6$ (MN) \cite{28} was prepared by 
calcining a stoichiometric mixture of $MgCO_3 .3H_2O$ and $Nb_2O_5$ at $1050 ^oC$ for six hours. At the 
next stage, stoichiometric amount of $TiO_2$ was mixed with $MgNb_2O_6$ and the mixture was calcined 
at $1050 ^oC$ for six hours to obtain $[(1-x)/3]MgNb_2O_6-(x)TiO_2$ (MNT) precursor. This MNT precursor 
was then mixed with stoichiometric amount of $PbCO_3$ and calcined at $750 ^oC$ for six hours. The 
powder obtained at this stage consists of pure perovskite phase of PMN-$x$PT free from the 
pyrochlore phase. Cold compaction of calcined powders was done using a steel die of 12-mm 
diameter and an uniaxial hydraulic press at an optimised load of 65 kN.  2 \% polyvinyl alcohol 
(PVA) solution in water was used as binder. The green pellets were kept at $500 ^oC$ for 10 hours 
to burn off the binder material and then sintered at $1150^oC$ for 6 hours in sealed crucibles with 
controlled PbO atmosphere. Density of the sintered pellets was higher than 98\% of the theoretical 
density. Sintered pellets were crushed into fine powders and then annealed at $500 ^oC$ for 10 hours 
to remove the strains introduced during crushing for x-ray characterizations. XRD measurements 
were carried out using a 12kW rotating anode (Cu) based Rigaku powder diffractometer operating 
in the Bragg-Brentano geometry and fitted with a graphite monochromator in the diffracted beam. 
Fired-on silver paste was used for electroding the sintered pellets. The dielectric measurements 
at 1kHz were carried out using a HIOKI 3532 LCR HiTester.

\section{Details of the Rietveld refinement}

Rietveld refinement was carried out using DBWS-9411 programme \cite{30}. In all the refinements 
pseudo-Voigt function was used to define the peak profiles while a fifth order polynomial was 
used for describing the background. Except for the occupancy parameters of the ions, which were 
kept fixed at the nominal composition, all other parameters i.e., scale factor, zero correction, 
background, half width parameters along with mixing parameters, lattice parameters, positional 
coordinates and isotropic thermal parameters were refined. The isotropic thermal parameter 
values for Pb was invariably found to be high as reported by Kiat et al also \cite{17}. Use of 
anisotropic thermal parameters for Pb did not lead to any improvement in the agreement factors. 

In the tetragonal phase with P4mm space group, the $Pb^{2+}$ ion occupies 1(a) sites at $(0,0, z)$, 
$Ti^{4+} / Nb^{5+} / Mg^{2+}$ and $O_I^{2-}$  occupy 1(b) sites at $(1/2,1/2, z)$ and 
$O_{II}^{2-}$ occupy 2(c) sites at $(1/2,0,z)$. 
For the rhombohedral phase with R3mH space group, we used hexagonal axes with lattice 
parameters $a_H = b_H = \sqrt{2} a_R$ and $c_H = \sqrt{3} a_R$ where $a_R$ corresponds 
to the rhombohedral cell 
parameter. In the asymmetric unit of the structure of the rhombohedral phase with R3mH space 
group, $Pb^{2+}$ and $Nb^{5+}/Ti^{4+}/Mg^{2+}$ ions occupy 3(a) sites at $(0,0,z)$ and $O^{2-}$ 
at the 9(b) site at 
$(x,2x,z)$. In the monoclinic phase with space group Cm, there are four ions in the asymmetric 
unit with $Pb^{2+}, Ti^{4+}/Nb^{5+}/Mg^{2+}$ and $O_I^{2-}$ in 2(a) sites at $(x,0,z)$ and 
$O_{II}^{2-}$ in 4(b) sites at $(x,y,z)$. 
The asymmetric unit of the monoclinic phase with space group Pm has got five ions with $Pb^{2+}$ 
and $O_I^{2-}$ in 1(a) site at $(x,0,z)$, $Ti^{4+}/Nb^{5+}/Mg^{2+}, O_{II}^{2-}$ and $O_{III}^{2-}$
 in 1(b) sites at $(x,1/2,z)$. Following the established conventions, $Pb^{2+}$ was fixed at (0,0,0) for
 the tetragonal \cite{31} and monoclinic \cite{4,16}structures. Following Megaw and Darlington \cite{32}, the 
$z-$coordinate of $O^{2-}$ was fixed at 1/6 for the rhombohedral structure.
Additionally, space group Bmm2 was also taken into consideration for a few 
compositions. For this space group, $Pb^{2+}$ ion occupies 2(b) sites at $(1/2,1/2,z), 
Nb^{5+}/Ti^{4+}/Mg^{2+}$ ions occupy 2(a) sites at $(0,0,z), O_I^{2-}$ in 4(d) sites at $(x,0,z)$ 
and $O_{II}^{2-}$ in the 2(b) sites at $(0,1/2,z)$. $Nb^{5+}/Ti^{4+}/Mg^{2+}$ ions were kept at the 
origin (0,0,0) for the refinment \cite{33}.

\section{ Location of the morphotropic phase boundary}	

Lead magnesium niobate (PMN) is a relaxor ferroelectric with very high value of room temperature dielectric 
constant ($ \approx 12000$ in ceramic form) [34]. With the addition of lead titanate (PT), whose dielectric 
constant is very low $(<400)$\cite{3}, it is expected that the dielectric constant of the resulting solid 
solution will decrease in comparison to PMN with increasing PT content. Fig.1 shows the variation of 
room temperature dielectric constant $\epsilon'$ with composition $(x)$ for the PMN-$x$PT ceramics in the 
composition range $0.20 \leq x \leq 0.45$. As expected, dielectric constant decreases with increasing value 
of $x$ upto $x = 0.26$. However, from $x=0.27$, the $\epsilon'-x$ plot takes an upward trend 
marking the onset of the 
MPB region. This upward trend continues upto $x \approx 0.30$ and thereafter shows a plateau region for 
$0.30 < x < 0.34$. Kelly et al \cite{35} have also observed a similar plateau region but in poled PMN-$x$PT 
samples. The dielectric constant again shows an upward trend leading to a peak around $x =0.35$ 
and thereafter it decreases monotonically with increasing $x$. The results shown in this figure 
correspond to averaging over at least 5 samples for each composition. The sample to sample 
variation of dielectric constant for each composition was less than 1\% for various PMN-$x$PT samples. 

Fig.1 reveals the presence 
of four different regions. For correlating these regions with a corresponding change in the crystal 
structure as a function of composition (x), we present in Fig.2 the powder XRD profiles of the 200, 
220 and 222 pseudocubic reflections for various PMN-$x$PT composition. For the composition range 
$0.20 \leq x \leq 0.26$, the 200 is a singlet, 
while 220 and 222 are doublets with weaker reflections occurring on the lower 2$\theta$ side. This 
characterizes a rhombohedral phase that is stable for $x \leq 0.26$. For $x > 0.26$, the width of the 200 
profile increases eventually leading to an asymmetric tail on the higher 2$\theta$ side which has become 
quite apparent for $x =0.29$ and 0.30. As shown in the next section, Rietveld analysis of the XRD data 
reveals that the structure of the dominant phase in the composition range $0.26 < x < 0.31$ is monoclinic ($M_B$ 
type) with Cm space group. The nature of the 200 profile again changes around $x = 0.31$ leading 
to the appearance of a shoulder on the higher 2$\theta$ side which eventually becomes a distinct peak 
with increasing $x$ as can be seen from Fig. 2 for $0.31 \leq x \leq 0.34$. In this composition range, the 
dominant phase is monoclinic ($M_C$ type) with Pm space group \cite{16}. For $x \geq 0.35$, the profiles 
shown in Fig. 2 
exhibit further changes. In particular, 200 pseudocubic reflection splits into 002 and 200/020 with 
nearly 1:2 intensity ratio. Further, the shoulder/peak on the lower 2$\theta$ side of the 220 pseudocubic 
profile is replaced by a distinct peak on the higher 2$\theta$ side. In addition, the 222 profile becomes a 
singlet. All these features correspond to the tetragonal structure and hence the dominant phase 
for $x \geq 0.35$ is tetragonal, as confirmed by the Rietveld analysis also, the results of which are 
presented in the next section. Thus the different regions shown in Fig. 1 correspond to four different 
crystallographic phases of PMN-$x$PT, which are stable over different range of composition.

\section{Rietveld analysis of XRD data}

\subsection{Rhombohedral structure with space group R3mH $(0 \leq x \leq0.26)$} 

Fig.3 depicts the observed, calculated and difference profiles obtained by Rietveld analysis of 
the XRD data for PMN-$x$PT with $x = 0.20$ and 0.26 using rhombohedral space group R3mH. The 
fit between the observed and calculated profiles is quite good confirming the rhombohedral 
structure of PMN-$x$PT for $x \leq 0.26$ in region I of Fig.1. The refined structural parameters and 
various agreement factors are given in Table1. 

\subsection{ Monoclinic structure with space group  Cm $(0.27 \leq x \leq 0.30)$}

For compositions with $x \geq 0.27$, the 200 reflection becomes broader
 which can not be accounted for in terms of the rhombohedral structure for which 200 is a singlet. 
This anomalous broadening is similar to that reported in $Pb(Fe_{1/2}Nb_{2/3})O_3$ \cite{36} and PZT with 
$0.530 \leq x \leq 0.62$ \cite{12}, where it has been attributed to the Cm phase. The anomalous broadening 
is absent for $x \leq 0.26$ as can be seen from the excellent fit shown in the insets (b) to Fig. 3 for the 
200 profile. To determine the true symmetry in the composition range $0.27 \leq x \leq 0.30$, we first 
carried out Rietveld refinements using various plausible space groups i.e., rhombohedral R3mH, 
monoclinic Cm, monoclinic Pm and orthorhombic Bmm2. Fig.4 shows the observed, calculated 
and difference profiles alongwith the various agreement factors for the pseudocubic 200, 220, 310 
and 222 reflections using four different space groups for $x = 0.29$. For the R3mH space group 
[Fig. 4(a)], we see that the mismatch between the observed and calculated profiles is quite 
prominent for 200 and 310 pseudocubic reflections, which is also confirmed by highest value 
of the agreement factors. Thus R3mH space group is simply ruled out. For the space group Pm, 
the misfit between the observed and calculated profiles for the 220 and 222 pseudocubic 
reflections is very large. In particular, for the 222 pseudocubic profile, the observed and calculated 
peaks are appearing at different 2$\theta$ values ruling out the possibility of the Pm phase. A similar misfit for 
the 220 and 222 profiles is observed for the Bmm2 space group also as can be seen from Fig. 4(c). 
The Cm space group gives the most satisfactory fit between the observed and calculated profiles 
for all the four reflections as can be seen from Fig.4 (d). This is corroborated by the lowest value 
of the agreement factors also. Fig. 5 depicts the observed, calculated and difference profiles in the 
2$\theta$ range 20 to 120 degrees for $x = 0.29$. The overall fit is quite satisfactory. The refined 
structural parameters are given in Table 2. From an analysis of the refined positional coordinates 
given in Table 2, it is found that the Cm phase of the PMN-$x$PT system is of $M_B$ type         
($Px = Py > Pz$) in contrast to the PZT system where the Cm phase corresponds to the $M_A$ type 
($Px = Py < Pz$) \cite{26}.

\subsection{ Monoclinic structure with space group Pm $(0.31 \leq x \leq 0.34)$}
	
On increasing the PT content beyond $x$ = 0.30, new features, like a shoulder in the 200 pseudocubic 
profile, appear. In order to determine the correct space group of PMN-$x$PT in this composition range, 
we considered Cm, Bmm2, and Pm space groups in our Rietveld analysis. Fig.6. depicts the observed, 
calculated and difference profiles of PMN-$x$PT with $x$ = 0.32 for the pseudocubic 200, 220 and 310 
reflections for the three space groups. It is evident from this figure that the best fit is obtained for the 
Pm space group which corresponds to the $M_C$ phase in the notation of Vanderbilt and Cohen \cite{26}. 
The agreement factors given in the last column of Fig. 6 are the lowest for the Pm space group. Fig.7 
depicts the observed, calculated and difference profiles in the 2$\theta$ range 20 to 120 degrees. 
The overall fit is quite satisfactory. Table 3 lists the refined structural parameters. It may be noted that the 
convention used for $\beta  (>90)$ in Table 2 is different from that $(<90)$used in reference \cite{16} as a result
of which the positional coordinates also appear to be different.

\subsection{ Tetragonal structure with space group P4mm $(0.35 \leq x \leq 1)$}	

Rietveld analysis for $x \geq 0.35$ confirmed that the dominant phase of PMN-$x$PT for these compositions 
has got tetragonal structure. Very good fit between the observed and calculated profiles were obtained 
using tetragonal P4mm space group as can be seen from Fig.8 for $x = 0.39$. The refined structural 
parameters are listed in Table 4 along with the agreement factors for this composition.

\subsection{Phase coexistence}	

One often observes coexistence of neighbouring phases in the MPB region due to extrinsic factors like 
compositional fluctuations \cite{37} and intrinsic factors like a first order phase transition between the low 
and high temperature phases \cite{13,15,38}. The results of the previous section show that in the PMN-$x$PT 
system, there are three phase boundaries occurring around $0.26 \leq x \leq 0.27$, $0.30 \leq x \leq 0.31$ and 
$0.34 \leq x \leq 0.35$ separating the stability fields of rhombohedral and Cm, Cm and Pm, and Pm and 
tetragonal phases, respectively. In order to see if further improvements in the agreement factors can 
result from a consideration of the coexistence of a minority neighbouring phase, we carried out Rietveld 
refinements for the composition ranges $0.27 \leq x \leq 0.30$, $0.31 \leq x \leq 0.34$ and  $0.35 \leq x \leq 0.39$
 using various plausible coexisting phases. It was found that in the composition range $0.27 \leq x \leq 0.30$, 
consideration of a minority rhombohedral phase led to higher agreement factors while minority
 monoclinic Pm phase decreased the agreement factors. The fits between observed and calculated
 profiles have improved for the (Cm + Pm) model as can be seen from a comparison of Fig. 4 (e) 
with Fig. 4 (d). Similarly, for the composition range $0.31 \leq x \leq 0.34$, consideration of a minority tetragonal
 phase decreased the agreement factors whereas the minority Cm phase increased the agreement 
factors. The presence of minority tetragonal phase improves the fit, especially on the lower 2$\theta$ side of the 
200 pseudocubic profile, as can be seen from a comparison 
of Fig.6 (c) with 6(d). Further, monoclinic Pm phase was found to coexist as a minority phase in the 
tetragonal region $0.35 \leq x \leq 0.39$ in agreement with the results of Ref. \cite{23}. The molar fractions of the 
minority and majority phases obtained by Rietveld refinement are plotted in Fig.9 as a function of PT content (x).
 It is evident from this figure that pure R3m phase exists for $x < 0.27$. For $x = 0.27$, the structure corresponds 
to that of pure Cm phase. On increasing the PT content (x), the Cm phase fraction decreases while the fraction of 
minority Pm phase increases.However, on crossing
 the Cm-Pm phase boundary at $0.30 < x < 0.31$, the fraction of the Pm phase increases abruptly. 
For $x = 0.31$, the structure corresponds to pure Pm phase. On increasing $x$ further ($>0.31$), 
the fraction of the majority Pm phase decreases while that of the minority P4mm phase increases with
 increasing x in the composition range $0.31 < x < 0.35$. For compositions with $x > 0.34$, the 
P4mm phase 
becomes the majority phase whose fraction increases with $x$ while the fraction of the minority Pm
phase continuously decreases.

\subsection{ Variation of lattice parameters with composition}	

Variation of lattice parameters with composition (x) for the majority phase is plotted in Fig. 10 for 
$0.20 \leq x \leq 0.45$. The [100] and [010] directions of the tetragonal phase correspond to the [010] and 
[100] directions of the Pm phase respectively. The [001] direction of the Pm phase deviates slightly 
from [001] direction of the tetragonal phase towards the [100] direction of the Pm phase giving 
rise to a monoclinic cell with unique b- axis. The [100] and [010] directions of the Cm phase, on the otherhand, are 
along the $<110>$ directions of the Pm and tetragonal phases. The cell parameters $a_m$, $b_m$ of the Cm 
phase are related to the elementary perovskite cell parameters 
$a_p$, $b_p$ as: $a_p \approx a_m$/$\sqrt2$ and $b_p \approx b_m$/$\sqrt2$. For the sake of easy 
comparison with the corresponding cell parameters of the tetragonal and Pm phases, we have plotted 
$a_p$ and $b_p$ instead of $a_m$, $b_m$ for the Cm phase in Fig.10. In order to maintain the polarization 
rotation path \cite{9,26} in going from tetragonal to Pm to Cm phases {for which $Pz \neq 0$ ($Px, Py = 0$), 
$Pz \neq Px \neq 0$
($Py = 0$) and $Px = Py \neq 0$, $Pz \neq 0$, respectively}, the $a, b, c$ axes of the Pm phase become 
$b_p$, $c_m$, $a_p$, respectively, of the Cm phase. It is evident from Fig.10 that for the tetragonal 
compositions, the $a-$ parameter increases while the $c-$ parameter decreases continuously with decreasing x.
 Around $0.34 < x < 0.35$, the $a-$ parameter of the tetragonal phase matches with the $b-$ parameter of the 
monoclinic (Pm) phase while the $c-$ parameter of the tetragonal phase, which remains as 
the $c-$ parameter of the Pm phase, shows a discontinuous drop. The $a-$ and $c-$ parameters of 
the Pm phase are nearly independent of composition but the b- parameter increases continuously 
with decreasing $x$. Further the monoclinic angle $\beta$ decreases continuously with decreasing x 
in the Pm phase field. The $b, a$ and $c$ parameters of the Pm phase, which become $c_m$, $b_p$ and 
$a_p$ of the Cm phase, do not show any discontinuity at the Pm-Cm phase boundary. Similarly, 
there is no discontinuous change in the $b_p$ and $c_m$ cell parameters at the Cm-R3m phase 
boundary but $a_p$ drops discontinuously. Table 5 lists the lattice parameter values of the 
majority phases for all the compositions studied by us.
\section{Concluding remarks}	
The phase diagram of Vanderbilt and Cohen \cite{26} for an eight order expansion of the free 
energy predicts the stability regions of three types of monoclinic phases, $M_A$, $M_B$, $M_C$, 
in addition to the tetragonal(T), rhombohedral(R) and orthorhombic(O) phases(see Fig.11). 
The R-$M_A$-T sequence of 
phase transition observed in PZT as a function of composition has been attributed to 
the region near $\alpha = \pi/2$, $\beta =0.102$ by Vanderbilt and Cohen. For the PMN-$x$PT system, we 
have shown that the stable phases in the composition ranges $x < 0.27$, $0.26 < x < 0.31$,
$0.30 < x < 0.35$ and $x > 0.34$ correspond to R, $M_B$, $M_C$ and 
T phases, respectively. In the phase diagram shown in Fig.11, the R- phase 
region is followed by a narrow stability region of the $M_B$ phase in broad agreement with our 
observations. However, as per this phase diagram, the $M_B$ and $M_C$ regions should be 
separated by a very thin orthorhombic (O) phase region. According to our Rietveld analysis results, 
the $M_B-M_C$  
phase boundary occurs around $0.30 < x < 0.31$. Interestingly, in the Rietveld refinement for $x = 0.30$, 
which is near this phase boundary, we found that $M_B$, $M_C$ and O phases give the same
 value of $R_{WP}$ (12.92) but the $R_B$ is 
the lowest for the O phase ($R_B$ = 12.84, 10.04, 9.92 for the $M_B$, $M_C$ and O phases, respectively)
 raising the possibility of the existence of the O phase in between the $M_B$ and $M_C$ phase regions.
 Thus the phase transition sequence $R - M_B - O -    M_C - T$ predicted by Vanderbilt and Cohen 
for $3\pi/4 < \beta < 0.8\pi$ may indeed correspond to the PMN-$x$PT system. Obviously, the structure of 
the PMN-$x$PT system in the MPB region is much more complex as compared to that in the PZT system 
with a simple  $R-M_A-T$ sequence of phase transitions. Although Vanderbilt and Cohen's theory predicts 
that $R-M_B$ phase boundary 
to be of first order type, our Rietveld analysis does not reveal any coexistence of R and 
$M_B$ phases. However, since the nature of the XRD profiles for the two phases are quite similar, except 
for the anomalous broadening of h00 and hh0 reflections for the $M_B$ phase, it may never be possible to settle
 the issue of coexistence of these two phases in a reliable fashion. The coexistence of $M_C$ 
and T phases revealed by our Rietveld analysis is not expected on the basis of the 
Vanderbilt and Cohen's theory since the corresponding boundary is of second order type. This 
could be due to the limitations of the eight order truncation of the free energy expansion. 

In the PZN-$x$PT system, there is some controversy 
about the structure of the MPB phase. According to Orauttapong et al \cite{17}, the sequence of 
phase transition in unpoled samples is R-O-T which is expected for $\pi < \beta < 3\pi/2$
 in the Vanderbilt and Cohen's phase diagram. Kiat et al have \cite{17}, however, shown that the 
structure of PZN-$x$PT in the MPB region for $x = 0.09$ corresponds to the $M_C$ phase. If it is so, 
we suspect the existence of $M_B$ phase and possibly O phase also interposed between R and 
$M_C$ phases, similar to what we have observed in the present study on the PMN-$x$PT system. It is
 likely that the relaxor ferroelectric based MPB systems may have similar sequence of phase 
transitions. Further, we suspect that the higher electromechanical response of these relaxor based MPB systems
may be linked with the ease of polarization rotation in the presence of $M_B$, $M_C$ and probably O phases
in the morphotropic phase boundary region as compared to the presence of only one phase ($M_A$) in the 
PZT system. 
\section{Acknowledgement}	
AKS acknowledges the award of senior research fellowship of UGC-BHU.

\newpage
\begin{center}
{\bf Figure Captions}
\end{center}

{\bf Fig.1} Variation of the real part of the dielectric constant ($\epsilon'$) with composition (x) at room temperature 
for PMN-$x$PT ceramics.

{\bf Fig.2} Evolution of the X-ray diffraction profiles of the 200, 220 and 222 pseudocubic reflections with 
composition (x) for PMN-$x$PT ceramics.

{\bf Fig.3} Observed (dots), calculated (continuous line) and difference (bottom line) profiles obtained after the 
Rietveld refinement of PMN-$x$PT with $x$=0.20 and $x$=0.26 using rhombohedral space group R3m
in the 2$\theta$ range 20 to 60 degrees. Inset (a) shows the patterns in the 2$\theta$ range 60 to 120 degrees
while the inset (b) illustrates the quality of fit for the 200 reflection. Tick marks above the difference profle show 
peak positions for CuK$\alpha1$.

{\bf Fig.4} Observed (dots), calculated (continuous line) and difference (bottom line) profiles 
of the 200, 220, 310 and 222 pseudocubic reflections obtained after the 
Rietveld refinement of PMN-$x$PT with x=0.29 using different structural models (a) Rhombohedral R3m
(b) Monoclinic Pm (c) Orthorhombic Bmm2 (d) Monoclinic Cm and (e) Monoclinic  (Pm+Cm) coexistence model.

{\bf Fig.5} Observed (dots), calculated (continuous line) and difference (bottom line) profiles obtained after the 
Rietveld refinement of PMN-$x$PT with $x$=0.29 using monoclinic space group Cm
in the 2$\theta$ range 20 to 60 degrees. Inset shows the patterns in the 2$\theta$ range 60 to 120 degrees.
Tick marks above the difference profle show peak positions for CuK$\alpha1$.

{\bf Fig.6} Observed (dots), calculated (continuous line) and difference (bottom line) profiles 
of the 200, 220, and 310 pseudocubic reflections obtained after the 
Rietveld refinement of PMN-$x$PT with x=0.32 using different structural models (a)  Monoclinic Cm
(b) Orthorhombic Bmm2 (c) Monoclinic Pm and (e) Monoclinic and tetragonal (Pm+P4mm) coexistence model.

{\bf Fig.7} Observed (dots), calculated (continuous line) and difference (bottom line) profiles obtained after the 
Rietveld refinement of PMN-$x$PT with for $x$=0.32 using monoclinic space group Pm
in the 2$\theta$ range 20 to 60 degrees. Inset shows the patterns in the 2$\theta$ range 60 to 120 degrees.
Tick marks above the difference profle show peak positions for CuK$\alpha1$.

{\bf Fig.8} Observed (dots), calculated (continuous line) and difference (bottom line) profiles obtained after the 
Rietveld refinement of PMN-$x$PT with $x$=0.39 using tetragonal space group P4mm
in the 2$\theta$ range 20 to 60 degrees. Inset shows the patterns in the 2$\theta$ range 60 to 120 degrees.
Tick marks above the difference profle show peak positions for CuK$\alpha1$.

{\bf Fig.9} Variation of molar fractions of different phases with composition (x) as obtained by Rietveld refinement.

{\bf Fig.10} Variation of lattice parameters with composition (x) for the majority phases of PMN-xPT.

{\bf Fig.11} Phase diagram for ferroelectric perovskites in the space of the dimensionless parameters $\alpha$
(vertical axis) and $\beta$ (horizontal axis) [after Ref. \cite{26}]
\newpage

\begin{center}

{\bf  Table 1.} Refined structural parameters of PMN-$x$PT for $x$=0.20 and 0.26 using rhombohedral space 
group R3mH.

\begin{tabular}{cccccccccccc}\hline
composition & & & & & Ions & & & & Positional coordinates & & Thermal parameters \\ \hline 
\end{tabular}

\begin{tabular}{ccccccccccccccccc}

(x) &  & & & & X & Y & Z & & & & B(${\AA}^2$) & & & \\ \hline 

0.20 & & & & $Pb^{2+}$ & 0.00 & 0.00 & 0.542(1) &&&&   3.02(1)   \\ 
0.26 & & & & $Pb^{2+}$ & 0.00 & 0.00 & 0.546(1) &&&&    2.91(2)  \\

0.20 & & & & $Ti^{4+}/Nb^{5+}/Mg^{2+}$ & 0.00 & 0.00 & 0.02(2) &&&& 0.61(7)  \\ 
0.26 & & & & $Ti^{4+}/Nb^{5+}/Mg^{2+}$ & 0.00 & 0.00 & 0.019(1)&&&&0.14(4)    \\ 

0.20 & & & & $O^{2-}$ & 0.353(3) & 0.176(3) & 0.1667 &&&&0.0(1)            \\ 
0.26 & & & & $O^{2-}$ & 0.325(3) & 0.162(3) & 0.1667 &&&&0.3(1)            \\ \hline
\end{tabular}
\begin{tabular}{cccccccccccccccccc} \\ 
0.20 &&$a=b=5.6921(1)$(${\AA}$)& &&$c= 6.9882(2)$  (${\AA}$)&&           \\
0.26 &&$a=b=5.6841(1)$(${\AA}$)& && $c= 6.9800(1)$  (${\AA}$)&&           \\ \hline
0.20& & $R_{WP}$= 14.76& $R_{exp}=6.46 $ & &$R_B =10.12 $& $\chi^2$ =5.06\\  
0.26&& $R_{WP}$= 12.98& $R_{exp}=7.45$ & &$R_B =9.93 $ &$\chi^2$ =3.01 \\ \hline 
\end{tabular}

\end{center}

\begin{center}

{\bf  Table 2.} Refined structural parameters of PMN-$x$PT for $x$=0.29 using monoclinic space group Cm.

\begin{tabular}{ccccccccccccc}\hline
&&& Ions & & & &&& Positional coordinates & & &Thermal parameters \\ \hline 
\end{tabular}

\begin{tabular}{ccccccccccccccccc}

& X & Y & Z & & & &B (${\AA}^2$) & & & \\ \hline 

$Pb^{2+}$ & 0.00 & 0.00 & 0.00& &&& 3.08(2)                 \\

$Ti^{4+}/Nb^{5+}/Mg^{2+}$ & 0.5250(8) & 0.00 & 0.498(2)&&&&0.73(4)    \\ 

$O_I^{2-}$ & 0.54(1) & 0.00 & -0.01(2)   &&&& 0.2(3)          \\ 
$O_{II}^{2-}$ & 0.317(2) & 0.267(4) & 0.48(1) &&&& 0.3(2)            \\ \hline
\end{tabular}
\begin{tabular}{cccccccccccccccc} \\ 
$a=5.6951(2)$(${\AA}$)& &$b=5.6813(2)$(${\AA}$)&&$c=4.0138(1)$(${\AA}$)& &$\beta =90.136(3)(^0)$  \\ \hline
$R_{WP}$= 12.24&&$R_{exp}=6.46 $ & &$R_B =7.10 $& & $\chi^2$ =3.59\\ \hline 
\end{tabular}

\end{center}

\begin{center}
\newpage
{\bf  Table 3.} Refined structural parameters of PMN-$x$PT for $x$=0.32 using monoclinic space group Pm.

\begin{tabular}{ccccccccccccc}\hline
&&& Ions & & & &&& Positional coordinates & & &Thermal parameters \\ \hline 
\end{tabular}

\begin{tabular}{ccccccccccccccccc}

& X & Y & Z & & & &B (${\AA}^2$) & & & \\ \hline 

$Pb^{2+}$ & 0.00 & 0.00 & 0.00  &&&& 3.28          \\

$Ti^{4+}/Nb^{5+}/Mg^{2+}$ & 0.509(2) & 0.50 & 0.5479(7)&&&&0.20(3)    \\ 

$O_I^{2-}$ & 0.47(1) & 0.00 & 0.57(1)   &&&&0.2(2)          \\ 
$O_{II}^{2-}$ & 0.417(8) & 0.50 & 0.059(6) &&&& 0.2(2)            \\ 
$O_{III}^{2-}$ & -0.02(1) & 0.50 & 0.57(1) &&&& 0.0(3)            \\ \hline
\end{tabular}
\begin{tabular}{cccccccccccccccc} \\ 
$a=4.0183(2)$(${\AA}$)& &$b=4.0046(1)$(${\AA}$)&&$c=4.0276(2)$(${\AA}$)&$\beta =90.146(3)(^0)$ \\ \hline
$R_{WP}$= 10.63&&$R_{exp}=5.42 $ & &$R_B =9.56 $& $\chi^2$ =3.84\\ \hline 
\end{tabular}

\end{center}
\begin{center}
{\bf  Table 4.} Refined structural parameters of PMN-$x$PT for $x$=0.39 using tetragonal space group P4mm.

\begin{tabular}{ccccccccccccc}\hline
&&& Ions & & & &&& Positional coordinates & & &Thermal parameters \\ \hline 
\end{tabular}

\begin{tabular}{ccccccccccccccccc}

& X & Y & Z & & & &B (${\AA}^2$) & & & \\ \hline 

$Pb^{2+}$ & 0.00 & 0.00 & 0.00  &&&&     2.92(1)      \\

$Ti^{4+}/Nb^{5+}/Mg^{2+}$ & 0.50 & 0.50 & 0.532(1)&&&&0.76(4)    \\ 

$O_I^{2-}$ & 0.50 & 0.50 & 0.054(4)   &&&&0.8(3)          \\ 
$O_{II}^{2-}$ & 0.50 & 0.00 & 0.601(2) &&&& 0.4(2)            \\ \hline
\end{tabular}
\begin{tabular}{cccccccccccccccc} \\ 
$a=3.9920(0)$(${\AA}$)&&&$c= 4.0516(1)$(${\AA}$)  &&       \\ \hline
$R_{WP}$= 13.85&&& $R_{exp}=6.75 $ & &&$R_B =10.12 $& & $\chi^2$ =4.21\\ \hline 
\end{tabular}

\end{center}
\newpage
\begin{center}
{\bf  Table5.} Refined cell parameters of PMN-$x$PT for the majority phases in the composition range 
$0.20\leq x \leq0.45$.
\begin{tabular}{cccccccccccccccccccccccccccc}\hline
&&Composition&&&&&& &&&Cell parameters &&& &&&&&&&&&&&               \\ \hline 
\end{tabular}

\begin{tabular}{ccccccccccccccccccccccccc}
&&&&(x)&&&&&&a(${\AA}$)& & &b(${\AA}$) &&c(${\AA}$)& &$\beta$(deg.) & &  \\ \hline  
&&&&0.20 && &&&&5.6921(1)& & && & 6.9882(2)  &&           \\
&&&&0.26 && &&&&5.6841(1)& & && & 6.9800(1)  &&           \\
&&&&0.27 && &&&&5.7001(2)& & &5.6852(2)& & 4.0186(1)  & &90.126(3)       \\
&&&&0.28 && &&&&5.6975(2)& & &5.6814(2)& & 4.0159(2)  & &90.133(3) \\
&&&&0.29 && &&&&5.6953(2)& & &5.6813(2)& & 4.0138(1)  &&90.136(3)           \\
&&&&0.30 && &&&&5.6962(3)& & &5.6806(2)& & 4.0123(2)  &&90.131(3)       \\
&&&&0.31 && &&&&4.0193(2)& & &4.0082(2)& & 4.0288(2)  &&90.145(3)       \\
&&&&0.32 && &&&&4.0183(2)& & &4.0046(1)& & 4.0276(2)  &&90.146(3)       \\
&&&&0.33 && &&&&4.0185(2)& & &4.0026(1)& & 4.0274(1)  &&90.169(2)       \\
&&&&0.34 && &&&&4.0174(2)& & &4.0019(2)& & 4.0289(2)  &&90.177(3)       \\
&&&&0.35 && &&&&4.0004(1)& & &&& 4.0464(1)         \\
&&&&0.36 && &&&&3.9970(1)& & &&&4.0468(1)         \\
&&&&0.37 && &&&&3.9953(1)& & &&&4.0492(1)         \\
&&&&0.38 && &&&&3.9933(0)& & &&&4.0495(1)         \\
&&&&0.39 && &&&&3.9920(0)& & &&&4.0516(1)         \\
&&&&0.45 && &&&&3.9832(1)& & &&&4.0579(1)         \\
 \hline
\end{tabular}
\end{center}


\begin{thebibliography}{99}

\bibitem{1} S. -E. Park and T. R. Shrout $\bf{82}$, 1804 J. Appl. Phys. (1997).
\bibitem{2} D. Viehland, A. Amin and J. F. Li $\bf{79}$, 1006 Appl. Phys. Lett. (2001).
\bibitem{3} B. Jaffe, W.R. Cook and H. Jaffe, Piezoelectric Ceramics, Academic Press, 
(London/New York), (1971).
\bibitem{4} B. Noheda, J. A. Gonzalo, R. Guo, S. -E. Park, L. E. Cross, D. E. Cox and G. Shirane,  
Phys. Rev. B $\bf{61}$ 8687 (2000).
 \bibitem{5} B.Noheda, D. E. Cox, G. Shirane, R. Guo, B. Jones and L. E.  Cross Phys. Rev. B  
     $\bf{63}$ 014103-1 (2000).
\bibitem{6} Ragini, S. K. Mishra, D. Pandey, H. Lemmens and G. V. Tendeloo Phys. Rev. B $\bf{64}$ 054101-1
 (2001).
\bibitem{7} R. Ranjan, Ragini, S. K. Mishra and D. Pandey Phys.Rev.B. , (R) $\bf{65}$,   060102             (2001)
\bibitem{8} D.M. Hatch, H.T. Stokes, R.Ranjan , Ragini, S.K. Mishra, D.Pandey and      
B.J.Kennedy, Phys. Rev. B $\bf{65}$, 2121101 (2002).
\bibitem{9} H. Fu, and R. E. Cohen, Nature $\bf{403}$, 281 (2000).
\bibitem{10} L. Bellaiche, and D. Vanderbilt Phys. Rev. Lett. $\bf{83}$ 1347 (1999); L. Bellaiche, 
A. Gracia and D. Vanderbilt Phys. Rev. Lett. $\bf{84}$ 5427 (2000).
\bibitem{11} R. Guo, L. E. Cross, S. -E. Park, B. Noheda, D. E. Cox and G. Shirane Phys. Rev.   Lett. $\bf{84}$
 5423 (2000).
 \bibitem{12} Ragini, R. Ranjan, S. K. Mishra, and D. Pandey J.Appl.Phys. 2002 July 15 issue.
\bibitem{13} S. K. Mishra, A.P. Singh and D. Pandey, Phil. Mag. B: $\bf{76}$, 213 (1997), 
ibid      Phil Mag. B: $\bf{76}$, 227 (1997).
\bibitem{14} D. L. Corker, A. M. Glazer, R. W. Whatmore, A. Stallard and F. Fauth 
J. Phys.: Condens. Matter $\bf{10}$ 6251 (1998).
\bibitem{15} Ragini, R. Ranjan,  S. K. Mishra  and  D. Pandey, (Submitted to Phys. Rev. B, for publication).
\bibitem{16} A. K. Singh and D. Pandey, J. Phys. Condens. Matter $\bf{13}$, L931 (2001)
\bibitem{17} J. M. Kiat, Y. Yesu, B. Dkhil, M. Matsuda, C. Malibert and G. Calvarin, 
Phys. Rev. B $\bf{65}$, 064106 (2002).
\bibitem{18} Y. Lu, D.-Y. Jeong, Z.-Y. Cheng, Q.M. Zang, H.-S. Luo, Z.-W. Yin, and D. Viehland, 
Appl. Phys. Lett. $\bf{78}$, 3109 (2001).
\bibitem{19} Z. -G. Ye, B. Noheda, M. Dong, D. E. Cox and G. Shirane, Phys.Rev. B. $\bf{64}$, 184114 (2001).
\bibitem{20} G. Xu, H. Luo, H. Xu, and Z. Yin, Phys. Rev. B. $\bf{64}$, 020102 (2001).
\bibitem{21} D. La-Orauttapong, B. Noheda, Z. -G. Ye, P.M. Gehring, J. Toulouse, D.E. Cox and
G. Shirane Phys. Rev. B $\bf{65}$, 144101 (2002).
\bibitem{22} M. K. Derbin, J. C. Hicks, S. -E. Park and T. R. Shrout, J. Appl. Phys., $\bf{87}$, 8159 (2000).
\bibitem{23} B. Noheda, D. E. Cox, G. Shirane, Z.-G. Ye, and J. Gao, arXiv: 
cond-mat/0203422 v1 20 March (2002).
\bibitem{24} B. Noheda, D. E. Cox, G. Shirane, S. -E. Park, L. E. Cross and Z. Zhong, 
Phys. Rev. Lett. $\bf{86}$, 3891 (2001).
\bibitem{25} K. Ohwada, K. Hirota, P. Rehrig, P. M. Gehring, B. Noheda, Y. Fujii, S. -E. Park and  G. Shirane, 
J. Phys. Soc. Japan $\bf{70}$ 2778 (2001). 
\bibitem{26} D. Vanderbilt and M. H. Cohen Phys. Rev. B $\bf{63}$ 094108 (2001).
\bibitem{27} A. K. Singh and D. Pandey (To be published ).
\bibitem{28} S.L.Swartz and T.R.Shrout, Mater.Res.Bull., $\bf{17}$, 1245 (1982).
\bibitem{29} O. Bouquin, L. Martine, J. Am. Ceram. Soc. $\bf{74}$ [5] 1152 (1991); H. C. Wang and W. A. Schulze 
J. Am. Ceram. Soc. $\bf{73}$ [4] 825 (1990).
\bibitem{30} R. A. Young, A. Sakthivel, T. S. Moss and C. O. Paiva Santos, Program DBWS-9411 for
 Rietveld  Analysis of x-ray and Neutron  Powder  Diffraction  Pattern (1994).
\bibitem{31} A. M. Glazer and S. A. Mabud, Acta Crystallogr., Sect. B; Struct. Crystallogr. Cryst. Chem. {\bf 34} 1065
(1978)
\bibitem{32} H. D. Megaw and C. N. W. Darlington, Acta Crystallogr. {\bf A31} 161 (1975).
\bibitem{33} H. D. Megaw, Ferroelectricity in Crystals, Methuen \& Co. Ltd. , 36 Essex Street, Londan, WC2 (1957).
\bibitem{34} S. L. Swartz, T. R. Shrout, W. A. Schulze and L. E. Cross, J. Am. Ceram. Soc.   $\bf{67}$ [5] 311(1984).
\bibitem{35} J. Kelly, M. Leonard, C. Tantigate and A. Safari, J. Am. Ceram. Soc. $\bf{80}$ 957 (1997).
 \bibitem{36} N. Lampis, P. Sciau and A. G. Lehmann, J. Phys: Condens. Matter $\bf{11}$, 3489 (1999).
\bibitem{37} K. Kakegawa, J. Mohri, K. Takahasi, H. Yamamura and S. Shirasaki., Solid State Commun. 
$\bf{24}$, 769,  (1977). 
\bibitem{38} S. K. Mishra, A.P. Singh and D. Pandey, Appl. Phys. Lett. $\bf{69}$, 1707 (1996).

\end{thebibliography}
\end{document}